\documentclass[10pt,letterpaper]{article}
\usepackage{opex3}
\usepackage{color}
\usepackage{cite}

\begin{document}

%%%%%%%%%%%%%%%%%% title page information %%%%%%%%%%%%%%%%%%
\title{Quantum communication with macroscopically bright nonclassical states}

\author{Vladyslav C. Usenko,$^{1,*}$ Laszlo Ruppert$^1$ and Radim Filip$^{1}$}

\address{$^1$Department of Optics, Palacky University, 17. listopadu 12, 771 46 Olomouc, Czech Republic}

\email{$^*$usenko@optics.upol.cz} %% email address is required



\begin{abstract}
We analyze homodyne detection of macroscopically bright multimode nonclassical states of light and propose their application in quantum communication. We observe that the homodyne detection is sensitive to a mode-matching of the bright light to the highly intense local oscillator. Unmatched bright modes of light result in additional noise which technically limits  detection of Gaussian entanglement at macroscopic level. When the mode-matching is sufficient, we show that multimode quantum key distribution with bright beams is feasible. It finally merges the quantum communication  with classical optical technology of visible beams of light.  
\end{abstract}

 % REPLACE WITH CORRECT OCIS CODES FOR YOUR ARTICLE, MINIMUM OF TWO; Avoid using the OCIS codes for “General” or “General science” whenever possible.
% 270.0270   Quantum optics
% 060.5565   Quantum communications
% 270.5568   Quantum cryptography
% 270.5570   Quantum detectors
% 270.5585   Quantum information and processing
% 270.6570   Squeezed states

%%%%%%%%%%%%%%%%%%%%%%% References %%%%%%%%%%%%%%%%%%%%%%%%%
%\bibliographystyle{osajnl}
%\bibliography{optex}
%\begin{thebibliography}{99}
%\end{thebibliography}

%\bibitem{Cerf2007}
%N.~J. Cerf, G.~Leuchs, and E.~S. Polzik, {\it Quantum Information with Continuous Variables of Atoms and Light} (Imperial College Press, 2007).

%%%%%%%%%%%%%%%%%%%%%%%%%%  body  %%%%%%%%%%%%%%%%%%%%%%%%%%
\section{Introduction}

%Standard \LaTeX{} or AMS\TeX{} environments should be used to place tables, figures, and math. Examples are given below.

%\begin{verbatim}
%\begin{figure}[htbp]
%\centering\includegraphics[width=7cm]{opexfig1}
%\caption{Sample caption (Ref. \cite{Oron03}, Fig. 2).}
%\end{figure}

%\begin{equation}
%H = \frac{1}{2m}(p_x^2 + p_y^2) + \frac{1}{2} M{\Omega}^2
%     (x^2 + y^2) + \omega (x p_y - y p_x).
%\end{equation}
%\end{verbatim}

Quantum physics revealed many advantages overcoming the limits of
classical physics. While classical physics deals with large macroscopic systems,
quantum physics describes various non-classical effects in
microscopic systems. During the last decades, the control of microscopic quantum
systems, like photons and atoms \cite{Cerf2007}, allowed to test the potential of quantum
physics in communication, metrology and computing. In many cases the
control remains quite challenging because of sensitive and tiny character
of microscopic systems. The key
solution can be found in macroscopic quantum states of light which are
easy to handle like in classical optics, but still keep relevant quantum features, which are needed for a
given task. A macroscopic nature however simultaneously means that large number of atoms
or optical modes cannot be precisely controlled. It is contradictory to existence of many very sensitive quantum phenomena. A
loss of single particle or mode can often cause a large reduction of quantum
nonclassical effects, which could bring quantum states closer to classical ones.

As an example of the macroscopic nonclassical quantum states, bright twin beams and entangled states of large numbers of atoms have been generated during the last decades \cite{Hald1999,Fernholz2008,Iskhakov2009,Toth2010,Iskhakov2012,Behbood2014,Vasilakis2015}. The bright photonic twin beams represent an empirical
evidence of non-classical states visible by a naked eye, they are
therefore easy to manipulate like the classical light beams. The twin beams
are non-classical because they are highly correlated in energy, which
contradicts classical wave theory \cite{Vasilyev2000}. High energy of both beams arises
from large number of the modes, which results in the macroscopic character of the
beams. The photon number correlations are fortunately robust against a loss
of few modes or receiving of a small number of unmatched modes in
the detectors. Such robustness allows one to detect quantum features of
macroscopic beams. A single pair of correlated modes inside the twin beam is also entangled in
phase-sensitive continuous variables (CVs) of light \cite{Braunstein2005}. It can be detected by a
pair of single-mode homodyne detectors with strongly coherent local
oscillators (LO), serving as the phase references. Such entangled states have interestingly a direct application,
they can be used to generate a secure key between two distant
parties in CV quantum key distribution (QKD) \cite{Madsen2012}, which was previously also shown possible
with the coherent states of light \cite{Grosshans2003}.

The symmetrical multimode entangled states with low energy can still be
indistinguishably detected by multimode homodyne detectors, if the
sufficiently large-amplitude LOs are matching all the modes of entangled
states. The modes, which are not matched to the LO, are simply negligible, compared to the matched modes. 
Consequently, the multimode detectors allow to
generate a secure key equivalently to the single-mode detection as it was recently shown in \cite{Usenko2014}. Now as a final step towards the 
use of classical beams, in the sense of their macroscopic character, for quantum communication, we consider 
the bright multimode states and waive the assumption of the weakness of the beams used for quantum communication.

In this paper we show that as the energy of multimode beams increases, the homodyne detector starts to be
sensitive to the large incoming energy of the modes which are not matched to LO. This sensitivity in fact witnesses the macroscopically
large intensity received by a homodyne detector. When the macroscopic modes do not match any mode of the
LO, they directly contribute to the noise of the detector.
Homodyne detection then becomes more noisy because of the detection of macroscopic very bright
light. The homodyne measurements of weak states with additional unmatched signals was previously studied in the context of ultrafast detection \cite{Raymer1995}. In the current paper we show how the structure of the bright matched and unmatched signal and also how unbalancing of the homodyne detection influence the noise caused by the macroscopic character of the states, and then focus on the role of such noise in the detection of nonclassical properties of light as well as on the applicability of the macroscopic states to QKD. This type of noise can be small, however, it cannot be neglected since it comes
from outside of the detector. It can be fully exploited by an eavesdropper
to substantially reduce security of QKD. It therefore
represents the most relevant limitation for the application of macroscopic
nonclassical states of light in secure quantum communication.

Using the homodyne detection of bright nonclassical light, we 
propose the feasible CV QKD implementation with the macroscopically bright nonclassical beams. Our proposal opposes the discrete-variable QKD (see \cite{Gisin2002} for review),
based on the single photons and faint light beams, and represents the very natural extension of CV QKD on the macroscopic states.
We suggest an optimal way of implementing QKD with bright beams, which are
easy to handle since they can be visible by a naked eye. The homodyne detector
noise caused by the macroscopic beams is nonlinear, it depends on
intensity of impinging light. Due to this, a trade-off appears between
security and macroscopic nature when entanglement increases. A distance of
secure transmission is shorten by that noise, when intensity increases,
but the security can be still guaranteed for. It is therefore possible
to extend CV QKD to the macroscopic version.
In this way, QKD can be finally realized by visible
beams having the same features as the classical optical beams.

\section{Homodyne detection of bright multimode states}

%We consider the case, when the generally multimode bright source of nonclassical entangled states (twin beam also known as two mode squeezed vacuum) is measured by a single-mode homodyne detector. For simplicity 
In classical optics, light beams typically consist of large number of modes, which are not individually controllable and measurable. 
To demonstrate the delicate aspect of such macroscopic nature of the signal states we consider each of the beams consisting of $M$ modes matched with the respective LO modes (which are shared between the communicating parties or are generated locally \cite{Qi2015}, and serve as a phase reference), and $N$ unmatched modes which do not match any LO modes. We study the impact of the bright auxiliary modes, which are not matched by the LO modes, but can affect the quadrature measurement and contribute to the noise in the quadrature variance. The mode mismatch can be the result of the difference in the mode structures of the source and the LO, but it can also be imposed by the effects in the channel such as the mode dispersion, when some of the signal modes lose mutual coherence with the respective modes of the LO. 
%Thus, each of the trusted parties receives two modes and measures them with the single-mode homodyne detector based on the coupling of the signal to a single-mode local oscillator, shared between the trusted parties, . 
%We show that it has the pronounced effect on security of continuous variable quantum key distribution (CV QKD) in such a scheme and must be carefully taken into account in real-world implementation of the quadrature-based quantum communication with bright two-mode squeezed vacuum states. 
%
\begin{figure}
\centering
\includegraphics[width=0.5\textwidth]{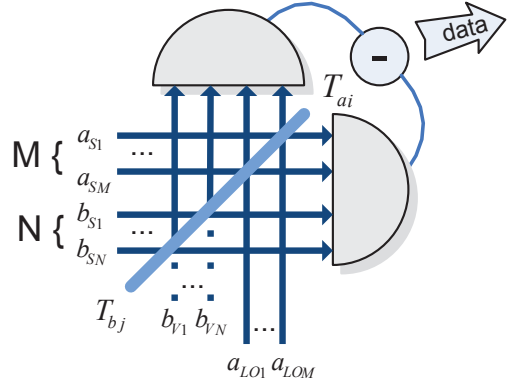}
\caption{Homodyne detection of macroscopic bright multimode states of light. Matched M signal modes $a_{S_i}$ are coupled to LO modes $a_{LO_i}$ and unmatched N signal modes $b_{S_j}$ are coupled to respective vacuum modes $b_{S_j}$ on a generally unbalanced beamsplitter with transmittances $T_{a_i}$ and $T_{b_j}$ for modes $a$ and $b$ respectively. The quadrature measurement consists in detecting the photocurrent difference between the two photodetectors on the output ports of the coupling beamsplitter.}
\label{Scheme_detection}
\end{figure}
In the multimode version of homodyne detection depicted in the Fig. \ref{Scheme_detection}, the coupling between the signal modes and the bright or vacuum modes of the LO modes is provided by a beamsplitter. One input of the beamsplitter is fed by the matched signal modes, characterized by the quantum operators $a_{S_1}..a_{S_M}$ and the auxiliary unmatched modes of the source, characterized by $b_{S_1}..b_{S_N}$. The other port is fed by the very bright LO modes $a_{LO_1}..a_{LO_M}$, being the classical phase reference for the homodyne detection, which are coupled to the modes $a_{S_i}$. The modes $b_{S_j}$ are coupled to the vacuum modes $b_{V_j}$, i.e., we assume that the modes $b_{S_j}$ are not matched by any of the bright modes of the LO. We thus assume here that the signal modes are either matched to the strong modes of the LO or not. Furthermore, if modes $b_{S_j}$ are weak, they do not play role in the detection since they are not amplified, but if they are strong they do play role and require no LO to be amplified. The coupling constants can be generally different for different modes entering the beamsplitter (i.e. the transmittance values of the coupling beamsplitter $T_{a_i}$ and $T_{b_j}$ are different for all the modes modes $a_{S_i}$ and $b_{S_j}$) and not exactly equal to $1/2$, which we refer to as an imperfect (unbalanced) homodyne detection. Note that vacuum input modes can also match the LO modes, but their effect would be negligible and is therefore not considered.
%
%\begin{equation}
%\left (
%\begin{array}{c}
% a_1 \\
% a_2
%\end{array}
%\right )
%=\left (
%\begin{array}{cc}
% \sqrt{T_a} & \sqrt{1-T_a} \\
%  -\sqrt{1-T_a} & \sqrt{T_a}
%\end{array}
%\right )
%\left (
%\begin{array}{c}
%a_S \\
%a_{LO}
%\end{array}
%\right )
%\end{equation}
%
%and similar for the quantum operators of the output modes $b_{\{1,2\}}$ and input modes $b_{\{S,0\}}$ thought the coupling $T_b$. 

The homodyne detection results in the photocurrent difference $\Delta_i$, which is proportional to the difference of the photon numbers $n_1-g\cdot n_2$ measured at the two mode-nondiscriminating detectors. Here $g$ is the coefficient applied during the measurement process to the data from one of the detectors, which is aimed at removing the contribution from the matched signal modes in the case of the unbalanced detection. This is the standard method in the homodyne detection and the role of the coefficient $g$ is discussed further in subsection \ref{ubdet}. Each of the photon numbers consists of the contributions from the output modes of the coupling beamsplitter $a^\prime_{S_i,LO_i}$ and $b^\prime_{S_j,V_j}$ (where ${*}^\prime$ indicates an output mode after the coupling), which are defined by the well known input-output relations of a beamsplitter:
\begin{eqnarray}
\label{inout}
\left( \begin{array}{c}
a^\prime_{S_i} \\
a^\prime_{LO_i}
\end{array} \right)_{out} =
\left( \begin{array}{cc}
\sqrt{T_{a_i}} & \sqrt{1-T_{a_i}} \\
-\sqrt{1-T_{a_i}} & \sqrt{T_{a_i}}
\end{array} \right)
\left( \begin{array}{c}
a_{S_i} \\
a_{LO_i}
\end{array} \right),
\end{eqnarray}
similarly for the conjugate operators $a^{\prime\dag}_{S_i},a^{\prime\dag}_{LO_i}$ defined through the $a^{\dag}_{S_i},a^{\dag}_{LO_i}$ and for the operators of output modes $b^\prime_{S_j},b^\prime_{V_j}$(and their conjugate), which are defined through the operators of the input modes $b_{S_j},b_{V_j}$ (and respectively their conjugate) and are governed by the coupling ratios $T_{b_j}$. 

The photon numbers detected by the photodetectors at the outputs of the coupling taking into account the suppression of the auxiliary modes then read $n_1=\sum_i^M{a^{\prime\dag}_{S_i}a^{\prime}_{S_i}}+\epsilon\sum_j^N{b^{\prime\dag}_{S_j}b^{\prime}_{S_j}}$ and $n_2=\sum_i^M{a^{\prime\dag}_{LO_i}a^{\prime}_{LO_i}}+\epsilon\sum_j^N{b^{\prime\dag}_{V_j}b^{\prime}_{V_j}}$. They can be obtained from the input-output relations Eq. (\ref{inout}) as
\begin{eqnarray}
\lefteqn{n_1=\sum_i^M\Big[T_{a_i}a^\dag_{S_i}a_{S_i}+\sqrt{T_{a_i}(1-T_{a_i})}\big(a^\dag_{S_i}a_{LO_i}+a^\dag_{LO_i}a_{S_i}\big)+(1-T_{a_i})a^\dag_{LO_i}a_{LO_i}\Big]+{}}
\nonumber \\
&&{}+\epsilon\sum_j^N\Big[T_{b_j}b^\dag_{S_j}b_{S_j}+\sqrt{T_{b_j}(1-T_{b_j})}\big(b^\dag_{S_j}b_{V_j}+b^\dag_{V_j}b_{S_j}\big)+(1-T_{b_j})b^\dag_{V_j}b_{B_j}\Big]
\end{eqnarray}
and
\begin{eqnarray}
\lefteqn{n_2=   \sum_i^M\Big[(1-T_{a_i})a^\dag_{S_i}a_{S_i}-\sqrt{T_{a_i}(1-T_{a_i})}\big(a^\dag_{S_i}a_{LO_i}+a^\dag_{LO_i}a_{S_i}\big)+T_{a_i}a^\dag_{LO_i}a_{LO_i}\Big]+{}}
\nonumber \\
&&{}+\epsilon\sum_j^N\Big[(1-T_{b_j})b^\dag_{S_j}b_{S_j}-\sqrt{T_{b_j}(1-T_{b_j})}\big(b^\dag_{S_j}b_{V_j}+b^\dag_{V_j}b_{S_j}\big)+T_{b_j}b^\dag_{V_j}b_{B_j}\Big].
\end{eqnarray}
Here we assume that the measurement devices are able to partially filter out the contribution from the unmatched modes. This can be done, for example, by frequency or spatial filtering, depending on the nature of the multimode structure of the macroscopic light. We characterize the inefficiency of mode filtration by the coefficient $\epsilon$, which is applied to the contribution from the auxiliary modes $b^\prime_{S_j},b^\prime_{V_j}$. Such inefficiency can vary in different detectors therefore in our analysis we consider it to be 
an unknown parameter of a particular set-up. Based on the estimation of $\epsilon$ for a given detector the possibility to filter out the unmatched signal can be estimated. We therefore use $\epsilon$ to characterize the robustness of the homodyne measurement of the macroscopic bright light and describe the ability of a homodyne detector to measure the macroscopic states on a microscopic level.

The LO in a $k$-th mode is a sufficiently strong coherent beam which can be approximated by a classical complex amplitude as $a_{LO_k}=\alpha\cdot e^{i\phi}$, where $\alpha$ is the real amplitude and $\phi$ defines the phase (both $\alpha$ and $\phi$ assumed to be the same for all the LO modes). Although the bright states are measured, the LO modes are still needed to be much brighter in order to reach effective and robust homodyne detection. Therefore the homodyne detection is measuring the observable proportional to $a_{S_k}^\dag e^{i\phi}+a_{S_k} e^{-i\phi}$ in the matched modes. If $\phi=0$, the homodyne detection measures the quadrature $x_k=a_{S_k}^\dag+a_{S_k}$ of the signal, while for $\phi=\pi/2$ the conjugate quadrature $P_k=i(a_{S_k}^\dag-a_{S_k})$ is measured. Further with no loss of generality we put $\phi$ to zero thus assuming that x-quadrature is measured in each of the signal modes.

\subsection{Balanced detection} 
We first consider the optimal and simplest case of the balanced detection, i.e. $T_{a_i}=T_{b_j}=1/2$ for any $i,j$ and $g=1$.  In this case the photon numbers read $n_1=\sum_i{a_{S_i}^{\prime\dag} a^\prime_{S_i}}+\epsilon \sum_j{b_{S_j}^{\prime\dag} b^\prime_{S_j}}$ and $n_2=\sum_i{a_{LO_i}^{\prime\dag} a^\prime_{LO_i}}+\epsilon \sum_j{b_{V_j}^{\prime\dag} b^\prime_{V_j}}$ and the photon-number difference expressed through the operators of the incoming modes taking into account the parametrization of the LO is
\begin{equation}
\Delta_i=\alpha\sum_i^Mx_i+\epsilon\sum_j^N{(b_{S_j}^\dag b_{V_j}+b_{V_j}^\dag b_{S_j})}.
\end{equation}
The measurement of the bright multimode light therefore consists of the balanced homodyne detection \cite{Collett1987} in the matched signal modes $a_{S_i}$ and the self-energy detection in the auxiliary signal modes $b_{S_j}$ resulting in the noise contribution. 

The noise arising from the brightness and multimode structure of the states becomes present in the quadrature variance measured by a homodyne detector. The quadrature variance is obtained from the variance of the difference photocurrent $Var(\Delta_i)=\langle \Delta_i^2 \rangle - \langle \Delta_i \rangle ^2$, where averaging is performed on the incoming states, namely the vacuum states in modes $b_{V_j}$ and thermal states (twin beams with one of the beams being traced out after measurement at Alice, or Gaussian-modulated squeezed states) in modes $a_{S_i}$ and $b_{S_j}$. The quadrature variance is essential when studying Gaussian nonclassical properties and their application in CV QKD. 
%explicitly characterized by the fist and second moments of quadratures, the latter are involved in the covariance matrix %shared between Alice and Bob, from which the entanglement and lower bound on the secure key can be derived. The noise %present in the state is defined in particular by the quadrature variance of the Bob's homodyne measurement on the %macroscopic multimode signal. 
Taking this into account and tracing out the vacuum modes $b_{S_j}$ we arrive at $\langle \Delta_i^2 \rangle = \alpha^2 \sum_i{\langle x_i^2 \rangle} + \epsilon^2 \sum_j{\langle b_{S_j}^\dag b_{S_j}\rangle} $ and $\langle \Delta_i \rangle = \alpha\sum_i\langle x_i\rangle$. The variance of the difference photocurrent must be normalized by the vacuum variance corresponding to the measurement taken with the input signal being blocked which is typically done in the experiment. For the detector being considered such variance would be $Var(\Delta_i^0)=M\alpha^2$, thus the normalized variance becomes
\begin{equation}
Var(\Delta_i)_{norm}=Var(X)+\epsilon_{tot}^2\bar{n},
\end{equation}
%
% where $\bar{n}\equiv\langle b_{S_j}^\dag b_{S_j} \rangle$ 
where $\bar{n}$ is the mean number of photons in a signal mode, $Var(X)$ is the quadrature variance of a signal mode, and 
\begin{equation}
\epsilon_{tot}^2=\frac{N\epsilon^2}{M\alpha^2}.
\end{equation}
Thus the quadrature variance even in the case of the perfectly balanced homodyne detection involves the noise term which is proportional to the mean photon-number in the additional signal modes. Clearly this inefficiency of the modes selection can be reduced by decreasing $N$ using better mode matching or by increasing amplitude $\alpha$ of the LO. By reducing the inefficiency $\epsilon_{tot}$ the noise term concerned with the macroscopic brightness of the signal can be strongly reduced. However it does not vanish as long as the bright unmatched modes are present. Therefore the homodyne detection of the multimode bright light becomes equivalent to the homodyne detection on the single-mode signal with the excess noise proportional to the mean photon number of a signal mode.

\subsection{Unbalanced detection} 
\label{ubdet}
If the homodyne detection of the macroscopic multimode signal is unbalanced, then the normalized photocurrent difference involves an additional term, proportional to the photon-number variance in the signal mode, $Var(n)$, so that in the two-mode case the normalized photocurrent difference reads
\begin{equation}
Var(\Delta_i)_{norm}^{(unb)}=Var(X)+\frac{\epsilon_{tot}^2}{T_a(1-T_a)}\Bigg[T_b(1-T_b)\bar{n}+(T_b-T_a)^2Var(n)\Bigg],
\end{equation}
taking into account that the coefficient $g$ applied to the contribution $n_2$ from the second detector is $g=\frac{T_a}{1-T_a}$ so that the term proportional to $a_s^{\dag}a_s$ vanishes. This way the impact of the photon number fluctuations in the main signal mode $a_s$ related to the unbalanced detection $T_a \ne 1/2$ is compensated, but the effect of the unbalancing remains present in the contribution from the unmatched modes.

In the non-macroscopic regime the noise related to the unmatched signal modes can be also present in the results of the homodyne detection, but its amount is very small due to the low mean photon number and therefore can be typically neglected especially if the intensity of the beam is additionally decreased by the channel attenuation. However in the macroscopic case, when the states are heavily multimode and intense, the noise arising from the macroscopic character must be taken into account as it can become crucial for transmission of the entanglement or for the secure key distribution as we show in the next Section. The limitations imposed by such the detection noise get more strict as the beams get closer to the macroscopic character. 
%In particular in the case of QKD as it originates from the uncontrolled auxiliary mode propagating though an untrusted channel and thus limits the security, which we discuss in the next Section. The limitation gets more strict as the beam approaches the macroscopic character. 

% (we assume for simplicity $N=M=1$)
%At the same time, the correlation between the measurements at Alice and Bob sides is scaled by $\sqrt{\frac{T_a}{1-T_a}}$. Therefore, the noise concerned with the macroscopic structure of the signal becomes even more pronounced in the unbalanced case. The additional term to the normalized photon-number variance proportional to $Var(n)$ has however minor impact on this variance compared to the term proportional to $\bar{n}$, when unbalancing is within few percents (it starts to dominate if unbalancing is more than 10\%, which is unlikely in a real experiment).

\section{Effects of macroscopic character on quantum resource and security}
\begin{figure}[h]
\centering
\begin{tabular}{lll}
\includegraphics[width=0.4\textwidth]{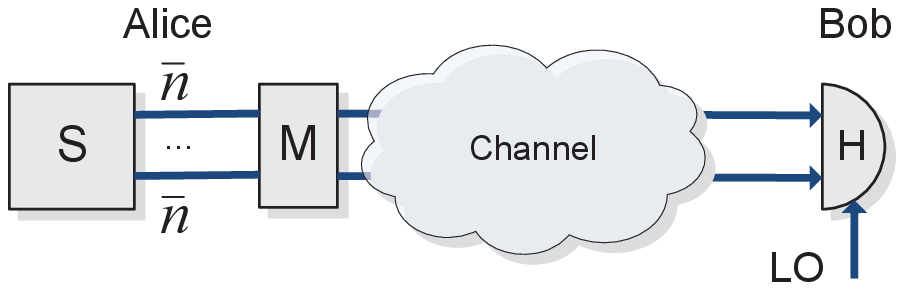}
\quad
\includegraphics[width=0.4\textwidth]{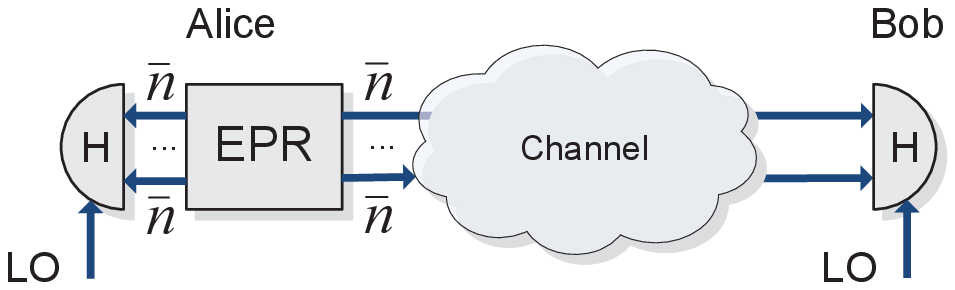}	
\end{tabular}
\caption{(Left): Prepare-and-measure and (Right): entanglement-based quantum communication schemes based on the multimode bright squeezed/entangled states and homodyne measurements.
\label{Scheme_communication}}
\end{figure}

We consider the effect of the noise concerned with the homodyning of macroscopic bright states on the detection of nonclassical resources contained in the states as well as on the application to secure CV QKD. To do so we consider two main quantum communication scenarios used to share a nonclassical resource or a secure key: prepare-and-measure (P\&M) and entanglement- (EPR-) based as depicted in the Fig. \ref{Scheme_communication}. In the case of P\&M scenario (Fig. \ref{Scheme_communication}, left) Alice possesses a source of the multimode bright squeezed light with quadrature variance $V_S<1$ in each mode, and applies modulation (random Gaussian-distributed displacement) to a signal up to the thermal state with variance $1/V_S$. The signal then travels through a quantum channel to a remote party Bob, who applies homodyne detection on the signal. The channel is characterized by transmittance $\eta$ and quadrature excess noise $\chi$, given in shot-noise units (SNU) of vacuum quadrature fluctuations. In the case of the EPR-based scenario (Fig. \ref{Scheme_communication}, right) Alice prepares the bright multimode entangled states, being the combination of twin beam states in each of the pairs of modes, and measures one of the beams by a local homodyne detector, while the other beam travels through the channel and is measured by Bob. We characterize the states by the mean photon number per mode $\bar{n}$ or by total mean photon number per beam $\bar{n}_{tot}=(M+N)\bar{n}$. 
%Equivalently, the twin-beam states $| \Psi \rangle = \frac{1}{\cosh{r}}\sum{\tanh{r}^n| nn \rangle}$ in two entangled modes can be characterized by the squeezing parameter $r$ so that the mean photon number and the photon-number variance in a signal mode are $\bar{n}=\sinh^2{r}$ and $Var(n)=(\sinh{r}\cosh{r})^2$ respectively. The noise concerned with the macroscopic character of the signal is therefore the monotonously increasing function of $r \in (0,\infty)$. 
%
The channel transforms a quadrature of a given i-th mode as $x_i^{'}=\sqrt{\eta}(x_i+x_N)+\sqrt{1-\eta}x_0$ (similarly for $p$-quadrature), where $x_N$ stands for the random displacement introduced by the channel excess noise with variance $Var(x_N)=\chi$, and $x_0$ represents a vacuum quadrature, corresponding to the pure channel loss, with $Var(x_0)=1$. Importantly, the channel transmittance degrades the signal by attenuating it, but also affects the noise arising from the macroscopic character of the signal, since the mean photon number $\bar{n}$ in a signal mode is scaled as $\eta\bar{n}$. This improves the applicability of the macroscopically bright nonclassical light in the practical communication scenarios over lossy channels. Further we consider the particular cases and the role of macroscopic structure of the states on the resources and security of CV QKD with such states.
\subsection{Homodyne detection of squeezing and entanglement} 
Before the detailed discussion of applicability of macroscopically bright light to secure quantum communication, we study the detection of basic nonclassical resources, contained in such light beams. It is easy to see, that the bright multimode quadrature squeezing $V_S<1$ with mean mode photon number $\bar{n}$ (defined by the squeezing and the possible displacement of a squeezed state) prepared by Alice and measured by the homodyne detector at the remote side in the P\&M scenario (Fig. \ref{Scheme_communication}, left) will be degraded by the noise concerned with macroscopic structure (contrary to a low energy case, when residual squeezing can be in principle observed upon any channel transmittance). The squeezing will become absent in the measurement independently of the channel transmittance when the mean photon number in a signal mode reaches $\bar{n}=(1-V_S)/\epsilon_{tot}^2$. For example, $V_S=-10$ dB of quadrature squeezing at $\epsilon_{tot}=10^{-2}$ would be undetectable at $\bar{n}\approx 10^4$ photons per mode.

\begin{figure}[h]
\centering
\begin{tabular}{ll}
\includegraphics[width=0.4\textwidth]{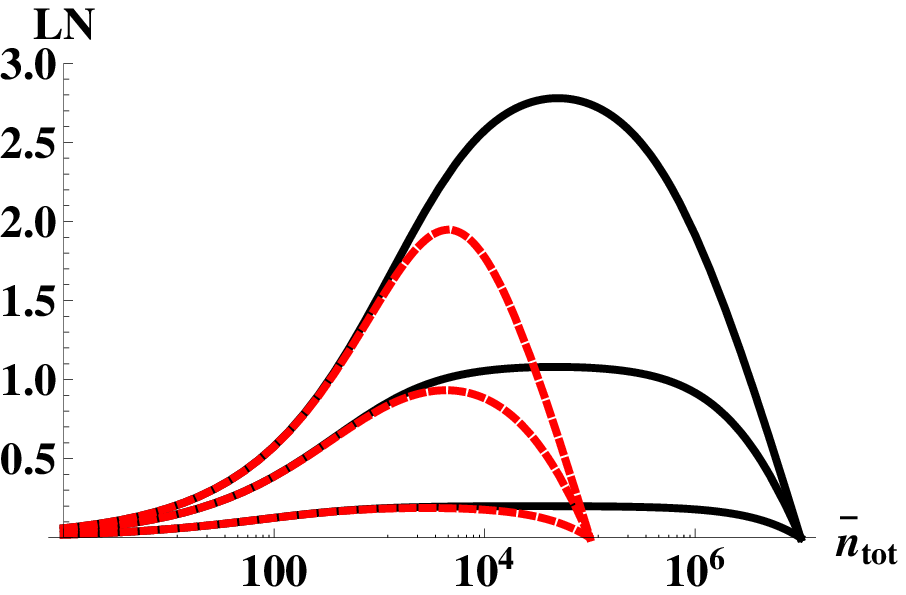}
\includegraphics[width=0.4\textwidth]{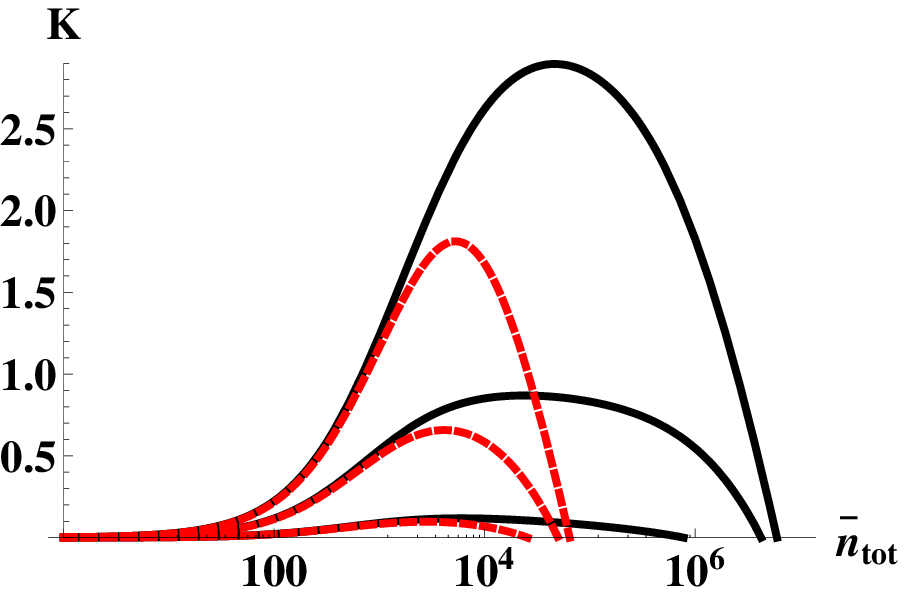}	
\end{tabular}
\caption{(Left): Logarithmic negativity of the macroscopic entangled states measured by the homodyne detectors versus total mean photon number. (Right): Key rate secure against collective attacks in bits per channel use, generated from the homodyne measurement of the macroscopic squeezed states versus total mean number of photons. On both the graphs the total number of modes is $10^3$ and 
$\epsilon_{tot}=10^{-2}$ (solid lines), $\epsilon_{tot}=0.1$ (red dashed lines). Channel transmittance (from bottom to top) $\eta=0.1,0.5,0.9$, channel noise is absent. 
\label{ENTKR}}
\end{figure}

Similarly the noise concerned with the homodyne detection of macroscopic states degrades the quantum entanglement resource of the states in the entanglement sharing scheme (Fig. \ref{Scheme_communication}, right). We analyze the effect of such noise on the entanglement resource in terms of the logarithmic negativity \cite{Vidal2002}, being the computable measure of entanglement. The entanglement is then lost independently of the pure channel loss ($\chi=0$) at $\bar{n}=\frac{1}{\epsilon_{tot}^2(1+\epsilon_{tot}^2/4)}$  being approximately $\bar{n}=\epsilon_{tot}^{-2}$. Therefore the proper suppression of the auxiliary modes can improve the detection of entanglement of the macroscopic states by the homodyne detectors. The dependence of the entanglement in the terms of the logarithmic negativity on the total mean photon number of the beams is given in. Fig. \ref{ENTKR}, left. It is evident from the plots, that for the given parameters of the detection and of the channel there is an optimal energy of the states that maximizes the entanglement.
%In particular, the entanglement is reduced already for a two-mode case and can completely vanish as illustrated in terms of the logarithmic negativity [] in Fig. \ref{entplot}. It is evident from the plot that the excess noise concerned with the macroscopic structure of the signal limits the entanglement of the states, which is more pronounced upon stronger channel loss. In the case of the balanced detector and perfect quantum channel ($\eta=1,\epsilon=0$) the entanglement is lost at 
%
%

%\begin{figure}[h]
%\begin{tabular}{ll}
%\includegraphics[width=0.22\textwidth]{ent2.eps}
%\includegraphics[width=0.22\textwidth]{ent_vs_n2.eps}	
%\end{tabular}
%\caption{(Left): Logarithmic negativity of the macroscopic entangled states measured by the homodyne detector versus weighting of the auxiliary modes for given photon number of a signal mode $\bar{n}=10^3$; (Right): Logarithmic negativity of the macroscopic entangled states measured by the homodyne detector versus mean number of photons in a signal mode and $\epsilon_{tot}=10^{-2}$. Solid lines: perfectly balanced detection; dashed and dotted lines: $T_b=1/2-1\%$; dotted line: $T_a=1/2+1\%$.
%\label{entplot}}
%\end{figure}
%
\subsection{Effect on security of CV QKD.} 
We consider the P\&M CV QKD scheme using macroscopic squeezed states as depicted in the Fig. \ref{Scheme_communication}, left. Such a scheme is equivalent to the EPR-based one \cite{Grosshans2003a} (Fig. \ref{Scheme_communication}, right) if the detection on the Alice's side is ideal (i.e., does not contain the noise arising from the macroscopic character), this equivalence will be used further. The security of CV QKD is analyzed using the covariance matrix formalism applied to the entangled state shared between the trusted parties, which is sufficient due to the extremality of Gaussian states \cite{Wolf2006} and subsequent optimality of Gaussian collective attacks \cite{Navascues2006,Garcia2006} in the assumption that an eavesdropper holds the purification of the untrusted noise. In our security analysis we therefore consider the effect of the macroscopic character of the signal on the covariance matrix of the state shared between the trusted parties. Since covariance matrices fully describe the Gaussian states, our approach complies with the optimal collective Gaussian attacks, while purification assumption allows us to analyze the security of the protocol not relying on a particular possible design of an eavesdropping attack. Note, however, that implementation-specific attacks on CV QKD \cite{Ma2013,Huang2013,Huang2014} (also referred to as quantum hacking) can be also possible in the macroscopic case. An eavesdropper can potentially explore e.g. the nonlinear response of the detectors in the strong energy regime, which would affect the calibration of the noise. Such attacks require a separate study based on the particular properties of the detectors. In the following study we focus on the general effects of the macroscopic character of the beams, which will unavoidably be present in any realization of CV QKD with macroscopic light and homodyne detection. 

The lower bound on the secure key in the case of collective attacks in the asymptotic regime and reverse reconciliation scenario (being stable against channel loss) is given by
\begin{equation}
K=\beta I_{AB}-\chi_{BE},
\label{krexp}
\end{equation}
where $\beta\in(0,1)$ is the post-processing efficiency (we further use $\beta=97\%$ following the efficiency of the existing error correcting codes \cite{Jouguet2011}), $I_{AB}$ is the mutual (Shannon) information between the trusted parties, and $\chi_{BE}$ is the Holevo bound, which upper limits the information leakage from the given channel. The information quantities involved in Eq. (\ref{krexp}) are obtained from the elements of covariance matrix of the state effectively shared between Alice and Bob \cite{Lodewyck2007}, taking into account the state and the channel parameters, and the noise arising from the macroscopic character of the beams. Such noise should be attributed to the channel (i.e., considered untrusted) since photon number in the auxiliary modes can be controlled by a potential eavesdropper who may thus hold the purification of the noise. Therefore, similarly to the distribution of entanglement, the trade-off between brightness or degree of entanglement of the source and the security of CV QKD appears. Indeed, the excess noise due to macroscopic structure can lead to the security break already in the case of a highly transmitting (even perfect) channel and perfectly balanced homodyne detection. Similarly to transmission of entanglement, it requires optimization of the total mean photon number for the given number of modes and $\epsilon_{tot}$, which effectively characterizes the imperfection of the homodyne detector, as shown in the Fig. \ref{ENTKR}, right, where the key rate has the similar profile as the Gaussian entanglement given in the Fig. \ref{ENTKR}, left (but contrary to entanglement the security is lost at different mean photon numbers for different transmittance values).
%Interestingly, there is an optimal two-mode squeezing of $r \approx 0.4$, which allows the protocol to tolerate the highest $\epsilon_{tot} \approx 0.35$ which is the maximum tolerable weight of the auxiliary signal photon-number contribution in the homodyne detection. 

\begin{figure}[h]
\centering
\begin{tabular}{ll}
\includegraphics[width=0.4\textwidth]{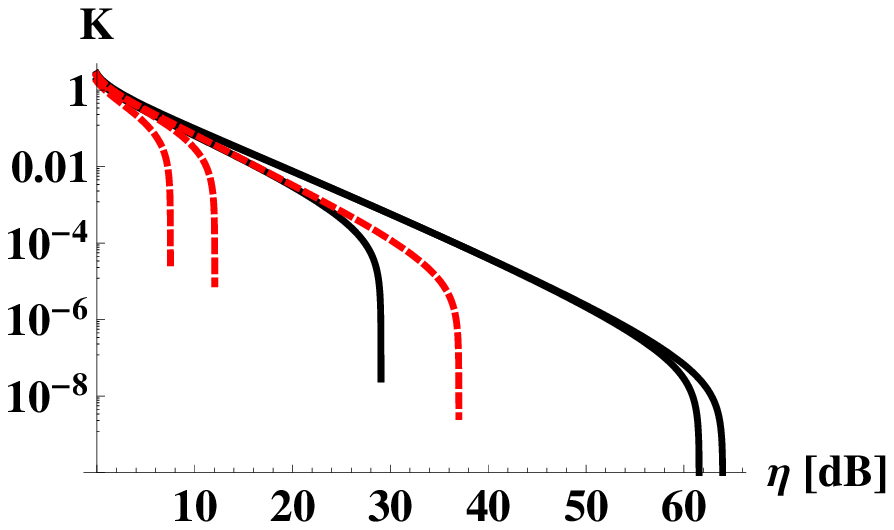}
\includegraphics[width=0.4\textwidth]{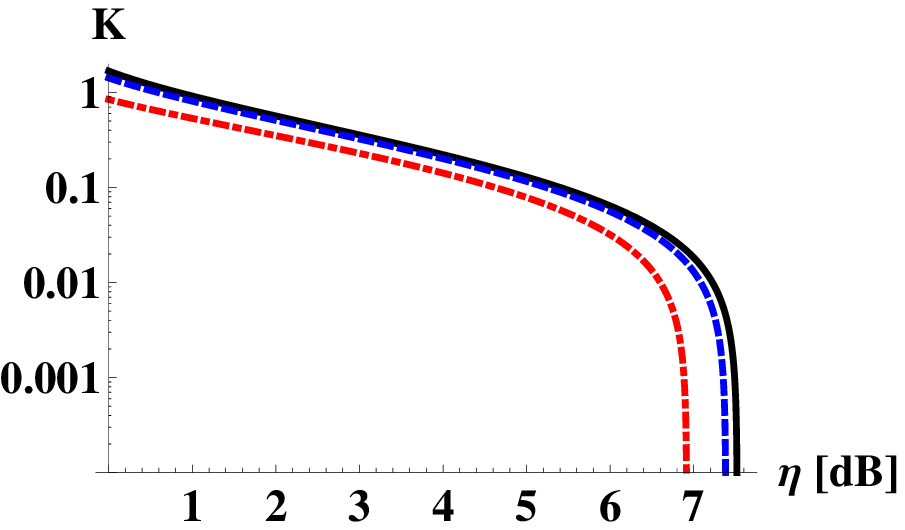}	
\end{tabular}
\caption{Key rate secure against collective attacks in bits per channel use, generated from the homodyne measurement of the macroscopic squeezed states plotted with respect to the channel transmittance $\eta$ (in negative dB scale) in the presence of channel noise $\chi=5\%$ SNU and upon $\epsilon_{tot}=0.1$. (Left): No detection unbalancing, perfect post-processing efficiency $\beta=1$ (black solid lines) and reduced efficiency $\beta=0.97$ (red dashed lines), $\bar{n}=10^3,10^2,10$ (from left to right). (Right): Reduced post-processing efficiency $\beta=0.97$, no unbalancing (black solid line), unbalanced detection of the signal $T_a=1/2+1\%$ (blue dashed line), additional unbalanced detection of the unmatched modes $T_b=1/2-1\%$ (red dot-dashed line).
\label{KR}}
\end{figure}

We show, that when the unmatched modes are properly suppressed by mode selection and/or power of the LO and when the number of modes is large (so that individually modes are not bright), the Gaussian CV QKD with bright multimode states is possible at the reasonably long distances, as depicted in the Fig. \ref{KR} (left), where the key rate versus channel attenuation for $\epsilon_{tot}=10^{-2}$ and for different mean photon numbers per mode is given. If the number of modes is high enough and the proper wavelength is chosen, such beams can be visible by a naked eye and become easier to handle. 

The unbalanced detection in the macroscopic regime slightly reduces the range of CV QKD with macroscopic states as shown in the Fig. \ref{KR}, right, but still enables the secure key distribution in the long-distance channels (the negative effect of the detection unbalancing becomes negligible for a better mode selection, i.e. lower $\epsilon_{tot}$ or for lower mean photon number per mode). Note that the additional noise arising from the generally unbalanced detection is scaled down by the square of the channel transmittance $\eta^2$, which reduces the negative effect. 

The limitation on the post-processing efficiency reduces the applicability of the protocol and requires stronger control over the number of modes (which need to be increased) or equivalently the mean photon number per mode (which must be reduced). Nevertheless, the CV QKD can be still deployed over strongly attenuating channels, corresponding to middle-range distances in a telecom fiber (with attenuation of $-0.2$ dB/km) or free space. This justifies the possibility to build CV QKD using experimentally generated bright macroscopic entangled states containing e.g. up to $10^5$ photons per pulse distributed by $10^4$ modes \cite{Iskhakov2012}. Assuming conversion to a telecom wavelength, presence of $5\%$ SNU of channel noise, realistic post-processing, and total auxiliary mode suppression inefficiency $\epsilon_{tot}=10^{-2}$, the CV QKD can be therefore implemented in the asymptotic limit at up to 180 km with $\bar{n}=10$, at up to 60 km with $\bar{n}=10^2$, and at up to 40 km with $\bar{n}=10^3$. The distances, however, would strongly depend on the particular implementation.
% with up to $\bar{n}=9\cdot10^3$ at the 10 km. distance (assuming conversion to a standard telecom wavelength and using the appropriate fiber) with suppression factor of $\epsilon_{tot} \approx 6.5\cdot 10^{-3}$ and on 50 km. distance with $\epsilon_{tot} \approx 2\cdot 10^{-3}$. For $\bar{n}=10^3$ the bounds would be weaker as $\epsilon_{tot} \approx 2\cdot 10^{-2}$ and $\epsilon_{tot} \approx 6\cdot 10^{-3}$ respectively and in the proper bandwidth such beams can be still visible by a naked eye. 
Note that if the bright multimode coherent states are used instead (i.e., $V_S=1$), the CV QKD can be still implemented, but the tolerable loss would be reduced by few dB. Therefore, with the strongly multimode states we can reach security using bright optical beams, which are easier to handle. Moreover, if the LO is reconstructed locally \cite{Qi2015}, the quantum signal appears to be the only state of light propagating through the channel and the brightness of the signal beams becomes crucial for handling the beams, which then waives the necessity in using additional bright pulses for beam pointing. Also the channel estimation with the stronger beams can be more efficient. This suggests the promising application of macroscopically bright nonclassical light in quantum communication.
\section{Conclusion} We shown that the impact of macroscopic character of light on the realistic homodyne detection of the macroscopically bright multimode signal results in the additional noise, which depends on photon numbers in the unmatched signal modes. Such noise reduces the entanglement and applicability of the states in secure quantum communication. However, if the mode-matching is efficient and/or brightness of the individual modes is reduced, the entanglement and squeezing can be observed by a homodyne detector. Based on this observation, we proposed the quantum communication scheme based on the macroscopic nonclassical light. We shown that QKD can be implemented with the macroscopically bright light. We therefore demonstrated the theoretical possibility of building secure quantum communication channels with the macroscopically bright nonclassical states of light.

\section*{Acknowledgments} The research leading the these results has received funding from the EU FP7 under Grant Agreement No. 308803 (project BRISQ2). V.C.U. and L.R.  acknowledge the project 13-27533J of GA\v CR.

\end{document}